\documentclass[reprint,prb]{revtex4-2}
\usepackage{graphicx}
\usepackage{physics}
\usepackage{color}
\usepackage{soul}

\newcommand{\up}{\uparrow}
\newcommand{\dn}{\downarrow}

\begin{document}

\title{Influence of leads on signatures of strongly-correlated zero-bias anomaly \\ in double quantum dot measurements}

\author{Caden Drover}
\author{R.L.\ Irvine} 
\author{Rachel Wortis}
\email{rwortis@trentu.ca}
\affiliation{Department of Physics \& Astronomy, Trent University, Peterborough, Ontario, K9L0G2, Canada}

\date{\today}

\begin{abstract}
The combination of disorder and interactions is known in many systems to produce a feature in the single-particle density of states, the shape and parameter dependence of which act as signatures of the underlying electronic state.
Strong Coulomb repulsion gives rise to a host of novel phenomena, among these is a unique zero-bias anomaly.
While understanding of the anomaly in bulk materials remains incomplete, a version of this anomaly can be found in an ensemble of two-site systems and hence has been predicted to be observable in parallel-coupled double quantum dots.
However, prior work did not include the influence of the leads.
Here we show that the presence of the leads results in changes to the projected stability diagrams but that the signature of the strongly-correlated zero-bias anomaly nonetheless remains clearly visible.
\end{abstract}

\maketitle 

\section{Introduction}
\label{intro}

Strong local electron-electron interactions lie at the heart of many interesting behaviours in transition metal oxide materials and other systems.  
A common probe of electron behaviour is the single-particle density of states (DOS), which has the benefit of being both readily measurable and calculable.  
The combination of interactions and disorder generically gives rise to a feature in the DOS which is pinned to the Fermi surface, and the shape and parameter dependence of this feature provide information on the underlying electronic state.  
Two well known examples are the Efros-Shklovskii Coulomb gap\cite{Efros1975} arising in insulating materials and the Altshuler-Aronov zero-bias anomaly\cite{Altshuler1985} arising in weakly disordered and weakly interacting systems.  
In strongly-correlated systems a disorder-induced suppression in the single-particle density of states has also been found,\cite{Chiesa2008,Song2009,Chen2011,Chen2012}   and the unique energy dependence makes it a potentially useful diagnostic of strong correlations.  
As a feature which arises due to an interplay of disorder, interactions, and mobility, it is referred to as a zero-bias anomaly, but its origin is distinct from that studied by Altshuler and Aronov.  
To emphasize this distinction we use the term strongly-correlated zero-bias anomaly (SZBA).  

In the limit of very strong disorder, a bulk system can be modelled as an ensemble of two-site systems.
Indeed an ensemble of two-site systems has a SZBA with the same dependence on parameters as that seen in numerical work on larger systems.\cite{Wortis2010,Chen2010,Wortis2011}
In particular, there is a regime in which the DOS is suppressed within a distance from the Fermi energy proportional to the hopping amplitude and independent of the strength of interactions or disorder.
Double quantum dots (DQDs) are physical two-site systems, and prior work proposed approaches to seeing the two-site SZBA in DQD measurements.\cite{Wortis2022}

Specifically, commonly measured stability diagrams, which plot the current through a double quantum dot as a function of the gate voltages on the two dots, display the physics of the two-site SZBA.  Most explicitly, an integral of the current over the stability diagram, plotted as a function of the bias voltage, corresponds to the disorder-averaged DOS as a function of energy, showing the opening of a SZBA as hopping between the dots is turned on.\cite{Wortis2022}
Moreover, the stability diagram itself displays the physics underlying this effect.
Two signatures stand out:  
First, inter-dot hopping transforms straight lines to curves, reducing the area of the stability diagram in which current flows as well as the magnitude of the current in some regions, corresponding to the suppression of the DOS of the ensemble at low energies.
Second, at higher bias voltage an abrupt increase in current occurs in specific regions of the stability diagram due to the entry of new transitions into the bias window, corresponding to the sharp increase in the density of states which marks the edge of the SZBA in the ensemble of two-site systems.

The calculations presented in Ref.\ \onlinecite{Wortis2022} assumed that current was carried by transitions exclusively from the ground state, neglecting the influence of the leads.
While this assumption is valid for the parameters specified, an obvious next question is whether these signatures of the two-site SZBA are still visible in DQD measurements when the system is not exclusively in its ground state at all times.
It is this question which we address here.
We find that the presence of the leads, treated in a weak-coupling approximation,
\footnote{The term zero-bias anomaly is also sometimes used to refer to features in the tunnelling conductance through mesoscopic devices associated with coherent tunnelling processes.  This is not the subject of this work.  The focus here is on the double dot system itself, and tunnelling is simply being used as a probe.} 
does result in changes to the stability diagram reported previously, but that signatures of the SZBA persist.
We also explore how these signatures depend on the relative strengths of the couplings to source and drain as well as on temperature.
This work demonstrates that evidence of the SZBA persists when coupling to the leads is taken into account and shows explicitly how it is eroded as the temperature increases, facilitating efforts to observe it in experiments.
Below we describe (Section \ref{Sec:Method}) the calculation of current within the weak-coupling approximation and then present (Section \ref{Sec:Results}) the resulting stability diagrams, comparing them with prior work.\cite{Wortis2022}

\section{Method}
\label{Sec:Method}

We consider a system consisting of a double quantum dot, leads, and tunneling between them.
\begin{eqnarray}
{\hat H} &=& {\hat H}_{dqd} + {\hat H}_{leads} + {\hat H}_{tunnel}
\end{eqnarray}
The quantum dot system consists of two dots with onsite Coulomb repulsion $U$, variable site potentials on each dot $V_i$, inter-dot Coulomb repulsion $V$, and inter-dot hopping $t_h$.
\begin{eqnarray}
{\hat H}_{dqd} &=& \sum_{j=1,2} (U {\hat n}_{j\up} {\hat n}_{j\dn} - e V_j {\hat n}_j)
				+ V {\hat n}_1 {\hat n}_1
				\nonumber \\ & &
				- t_h \sum_{\sigma= \up, \dn} ({\hat d}_{1\sigma}^{\dag} {\hat d}_{2\sigma} + {\hat d}_{2\sigma}^{\dag} {\hat d}_{1\sigma})
\end{eqnarray}
Here ${\hat d}_{j\sigma}^{\dag}$ and ${\hat d}_{j\sigma}$ are the fermionic creation and annihilation operators corresponding to site $j$ and spin $\sigma$; ${\hat n}_{j\sigma} \equiv {\hat d}_{j\sigma}^{\dag} {\hat d}_{j\sigma}$ is the spin-specific number operator on site $j$; and ${\hat n}_j \equiv \sum_{\sigma=\up,\dn} {\hat n}_{j\sigma}$ is the total number operator on site $j$.

This Hamiltonian has 16 eigenstates for which we establish the following notation:
the vacuum $\ket{0}$; 
two 1-particle bonding orbitals with spin $\sigma=\up,\dn$ $\ket{1b\sigma}$; two 1-particle antibonding orbitals $\ket{1a\sigma}$;
three 2-particle singlet states $\ket{2sa},\ket{2sb},\ket{2sc}$ (linear combinations of Fock states $\ket{20},\ \ket{02},$ and $(\ket{\up \dn} - \ket{\dn \up})/\sqrt{2}$ listed in order of increasing energy);
three 2-particle triplet states with spin $\tau=1,0,-1$ $\ket{2t\tau}$; 
two 3-particle bonding orbitals $\ket{3b\sigma}$; two 3-particle anti-bonding orbitals;
and the 4-particle state $\ket{4}=\ket{22}$.

The leads consist of source and drain, modeled as non-interacting Fermi seas. 
\begin{eqnarray}
{\hat H}_{leads} &=& \sum_{\ell, k, \sigma} \epsilon_{\ell k} {\hat c}_{\ell k \sigma}^{\dag} {\hat c}_{\ell k \sigma}
\end{eqnarray}
Here ${\hat c}_{\ell k \sigma}^{\dag}$ and ${\hat c}_{\ell k \sigma}$ are fermionic creation and annihilation operators corresponding to state $k$ and spin $\sigma$ in lead $\ell=s,d$, referring to source and drain, with spin-independent energy
$\epsilon_{\ell k}$.
We consider a parallel-coupled configuration such that 
hopping is allowed between both leads and both dots, with distinct hopping amplitude for the source $t_s$ and the drain $t_d$.
\begin{eqnarray}
{\hat H}_{tunnel} &=& - \sum_{j,\ell,k,\sigma} t_{\ell} ({\hat d}_{j\sigma}^{\dag} {\hat c}_{\ell k \sigma} + {\hat c}_{\ell k \sigma}^{\dag} {\hat d}_{j\sigma})
\end{eqnarray}

The effect of the leads is treated in the weak coupling (also known as sequential tunneling) limit.\cite{BruusFlensberg}
Specifically we assume that the lead hopping amplitude is sufficiently small that the time between tunneling events is long compared to the coherence time of the DQD, hence neglecting quantum interference between tunneling events.
In this limit, the state of the dot can be described by a classical rate equation, known as the Pauli equation.
The rate of change of the probability that the DQD is in the state $\alpha$, $P(\alpha)$, may be expressed in terms of a sum over rates of transition into and out of that state.
\begin{eqnarray}
{\partial \over \partial t} P(\alpha) &=& \sum_{\beta} ( \Gamma_{\beta \to \alpha} P(\beta) - \Gamma_{\alpha \to \beta} P(\alpha) )
\end{eqnarray}
The states $\alpha, \beta$ are eigenstates of ${\hat H}_{dqd}$, and the transition rates are calculated from Fermi's Golden Rule treating ${\hat H}_{tunnel}$ as a perturbation.  
\begin{eqnarray}
\Gamma_{\alpha \to \beta} &=& {2 \pi \over \hbar} |\bra{\beta,f} {\hat H}_{tunnel} \ket{\alpha,i}|^2 \delta(E_{\alpha,i}-E_{\beta,f})
\label{eqn:fgr}
\end{eqnarray}
Here $i$ and $f$ represent the initial and final states of the leads.

Considering each lead separately,
the rate of transition to a state $\beta\pm$ with $\pm 1$ particle relative to $\alpha$ through lead $\ell$ can be compactly expressed as
\begin{eqnarray}
\Gamma_{\alpha \to \beta\pm}^{\ell} &=& {2 \pi \over \hbar} |t_{\ell}|^2 \rho \ C_{\alpha \beta} n_F(E_{\beta} - E_{\alpha} \mp \mu_{\ell}) 
\label{eqn:gamma}
\end{eqnarray}
Here $\rho$ is the single-particle density of states in the lead, assumed flat within $k_B T$ of the chemical potential in the lead, $\mu_{\ell}$.  $n_F(x)=1/(1+e^{x/k_B T})$ is the standard Fermi function.  $C_{\alpha \beta}$ are coherence factors specific to the two states involved.  These are summarized in Appendix \ref{app:cf}.
The total rate $\Gamma_{\alpha \to \beta}= \sum_{\ell=s,d} \Gamma_{\alpha \to \beta}^{\ell}$.

We're interested in the stead-state behaviour of the dot, and hence $\partial P(\alpha)/\partial t=0$.  
We therefore wish to solve the equation $\underline{\underline \Gamma} {\vec P} = 0$ where 
$(\underline{\underline \Gamma})_{\alpha \beta} = \Gamma_{\alpha \to \beta}$, 
$(\underline{\underline \Gamma})_{\alpha \alpha} = - \sum_{\beta \ne \alpha} \Gamma_{\alpha \to \beta}$,
and $({\vec P})_{\alpha} = P(\alpha)$.  $\underline{\underline \Gamma}$ as constructed is indeed a singular matrix, allowing for nonzero ${\vec P}$.  In particular, since the elements of ${\vec P}$ are probabilities, we impose the normalization $\sum_{\alpha} P(\alpha)=1$.  
Replacing the last row of $\underline{\underline \Gamma}$ with this condition, we obtain the modified equation
\begin{eqnarray}
\underline{\underline \Gamma}' {\vec P} &=& {\vec V} 
\label{eqn:GPV}
\end{eqnarray}
where ${\vec V}=(0,0,0,...,1)$.  The result is independent of which row in $\underline{\underline \Gamma}$ is replaced.
Eqn.\ (\ref{eqn:GPV}) is then solved numerically using the python library NumPy.

This calculation results in a list of the occupation probabilities of each state of the DQD, which together with the transition rates allow the calculation of the steady-state current through the dot:
\begin{eqnarray}
I &=& \sum_{\alpha} P(\alpha) \left( \sum_{\beta+} \Gamma_{\alpha \to \beta+}^s - \sum_{\beta-} \Gamma_{\alpha \to \beta-}^s \right)
\end{eqnarray}
Here we present results in two formats.
First, stability diagrams, namely plots of the current as a function of the gate voltages, $V_1$ and $V_2$, on the two dots at a fixed value of the bias voltage $V_{sd}=\mu_s-\mu_d$.
Second, plots of the integrated current versus bias voltage, where the integrated current is the magnitude of the current summed over all points in the stability diagram (using a fixed range of $V_1$ and $V_2$). 

We use an asymmetric bias in which the chemical potential of the drain is held fixed at $\mu_d=(U/2)+V$, corresponding to half-filling when $V_1=V_2=0$, while $\mu_s=\mu_d+V_{sd}$. 
We present data for two values of the ratio of the strengths of coupling to source and to drain: $t_d/t_s=100$ and $t_d/t_s=1$.  
In both cases, $t_d$ and $t_s$ remain small enough that the states in the dot are not perturbed and coherence between tunnelling events can be neglected.
We focus first on the zero temperature limit and later explore the influence of nonzero temperature $T$ on our results.
We assume each dot is sufficiently small that the orbital level spacing is much larger than the thermal energy $k_B T$ and the bias energy $eV_{sd}$, such that only occupancies 0, 1, and 2 are relevant and all higher orbitals may be neglected.
The range of energies over which the dot potentials $eV_1$ and $eV_2$ are varied is larger than the onsite Coulomb repulsion $U$ which is in turn larger than the inter-dot hopping amplitude $t_h$.

\section{Results}
\label{Sec:Results}

In presenting our results we aim to address the following four questions:
First, is the stability diagram of the DQD different when leads are included as compared with the DOS approach used in Ref.\ \onlinecite{Wortis2022}? 
Second, is the SZBA still visible in the integrated current as a function of bias voltage? 
Third, having initially assumed stronger coupling to the drain than to the source, consistent with Ref.\ \onlinecite{Wortis2022}, how are the results affected when $t_d=t_s$?
And finally, how are the results affected by nonzero temperature?
We address each of these below.

\begin{figure}
\includegraphics[width=\columnwidth]{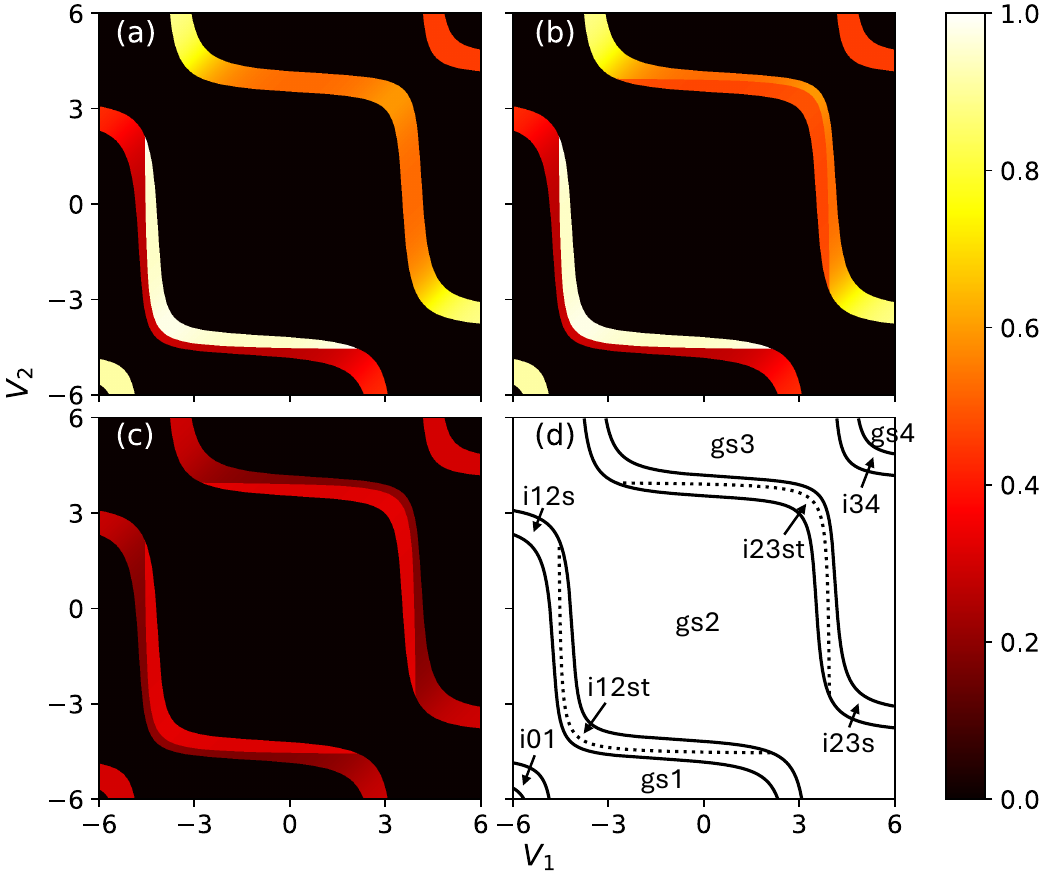}
\caption{\label{Fig:2dots_sd}
Stability diagrams for a parallel-coupled double quantum dot calculated using the DOS approach\cite{Wortis2022} (a) and the Pauli equation approach described in the text with (b) $t_d/t_s=100$ and (c) $t_d/t_s=1$.  Horizontal and vertical axes are the potential energy on the two dots, and the colour scale indicates the magnitude of the current.  All panels are at zero temperature with dot parameters $U=8,\ V=0.6,\ t_h=0.6$ and lead parameters $\mu_d=V+U/2,\ \mu_s=\mu_d+V_{sd},\ V_{sd}=0.6,\ t_s=0.0001$.
Panel (d) labels regions in the stability diagram for reference in the text.  Each region without current is labeled gs$n$ indicating that the ground state has $n$ particles in the region.  Each region in which current flows is labeled i$\alpha\beta$ where $\alpha$ and $\beta$ indicate the transitions involved as detailed further in the text.
}
\end{figure}

Consider first the question of whether the stability diagram produced by the Pauli method is different from that produced by the DOS approach.
Fig.\ \ref{Fig:2dots_sd}(a) shows the stability diagram produced by the DOS approach.  The content of this figure is the same as Ref.\ \onlinecite{Wortis2022} Fig.\ 4(d), while the colour scale in this paper has a smaller maximum value than that used in Ref.\ \onlinecite{Wortis2022} such that the features of interest are more clearly distinguished.  We briefly review the explanation of this figure:
In the DOS approach, current arises from single-particle transitions from a ground state set by the drain, $\mu_d$. 
Table \ref{tbl:2dotregions} (second column) lists the ground states in each of the regions of current labeled in Fig.\ \ref{Fig:2dots_sd}(d).
Starting at the lower left, there is current at the boundary, i01, between regions with no current (black) characterized by 0-particle occupancy (gs0) and 1-particle occupancy (gs1).
The current here arises from transitions between the vacuum state and the two 1-particle bonding orbitals: $\ket{0} \to \ket{1b\sigma}$.
In region i12s, the current arises from the transition $\ket{1b\sigma} \to \ket{2sa}$.
Region i12st has more current than region i12s due to the entrance of triplet states into the bias window.  
In both regions i23st and i23s, using the DOS approach, the current arises from the transitions $\ket{2sa} \to \ket{3b\sigma}$.
And finally in region i34, the current comes from $\ket{3b\sigma} \to \ket{4}$.  

Two questions arise:
First, why is there less current in i12s than in the i23 regions?  This is because in i12s current can flow through only one spin species (whichever one is not occupied in the 1-particle initial state), whereas in the i23 regions the ground state $\ket{2sa}$ allows either spin species to enter the dot.
Second, in the i23 regions, why is there less current closer to $V_1=V_2$ and more when they are more imbalanced?  This arises from the coherence factors $C_{\alpha \beta}$ in transition rates, Eqn.\ \ref{eqn:gamma}, and similar variation in shade is seen in i12s.

With this as context we can turn to Fig.\ \ref{Fig:2dots_sd}(b), which shows the stability diagram produced by the Pauli approach with $t_d/t_s=100$.
Much of the diagram is identical to Fig.\ \ref{Fig:2dots_sd}(a), with the same transitions responsible for the current.
The key distinction is seen in region i23st, where the current is lower in the Pauli approach than in the DOS approach.  We now examine the origin of this difference.

In the Pauli approach, no ground state is defined for the dot.
Instead, current is the net result of transitions from a set of initial states.
These states are determined by the energy differences between states $\Delta_{\alpha \to \beta}\equiv E_{\beta} - E_{\alpha}$.
Specifically, when $t_d/t_s=100$, the initial states are dictated by where the energy differences sit relative to the chemical potential of the drain, $\mu_d$.
Details, and a list of the initial states and their occupation probabilities, are shown in Appendix \ref{app:occup}.  
The transitions, single particle addition and removal, which allow the flow of current are those for which the transition energy falls in the bias window,
$\mu_d < \Delta_{\alpha \to \beta} < \mu_s$.

Consider first the region i23s. 
Here, the transition energies $\Delta_{0 \to 1b}, \ \Delta_{1b \to 2sa},$ and $\Delta_{1b \to 2t\tau}$ all fall below $\mu_d$.
It might therefore appear that $2sa$ as well as all three triplet states should have nonzero probability of occupancy.
However, $\mu_d < \Delta_{2sa \to 3b} < \mu_s$ while $\Delta_{2t\tau \to 3b}<\mu_d$.
Therefore, a system which starts in a triplet state will immediately transition to the 3-particle bonding state.
Once there it cannot transition back to a $\ket{2t\tau}$ state but only to the $\ket{2sa}$ state because only the latter transition falls in the bias window.  
Therefore, in steady state, $P(2sa)=1$ and the current is the same as that found using the DOS approach.

In the region i23st, the key distinction is that both $\Delta_{2sa \to 3b}$ and $\Delta_{2t\tau \to 3b}$ 
fall in the bias window.  
Therefore, $P(2sa)=P(2t,+1)=P(2t,0)=P(2t,-1)$.  
This results in less current than when $P(2sa)=1$ because either spin can flow through $\ket{2sa}$, but only one spin direction can flow through $\ket{2t,+1}$ and $\ket{2t,-1}$.  
The net result is a dark shadow at the 2-particle to 3-particle boundary which mirrors the bright sliver at the 1-particle to 2-particle boundary.

Note that there is no limit in which the results of the DOS and Pauli approaches match.  
This is in contrast to the stability diagram of a single dot. 
In the double-dot case, specifically in the regime of 2-particle occupancy, not only are there multiple initial states but crucially they do not all have equal energy ($E_{2sa} \ne E_{2t\tau}$) and this does not arise in the single-dot case.
When all initial states do have the same energy, then in the limit of large $t_d/t_s$ the two approaches produce the same results.
This is the case for the single dot and also for the double dot with zero inter-dot hopping.

\begin{figure}
\includegraphics[width=\columnwidth]{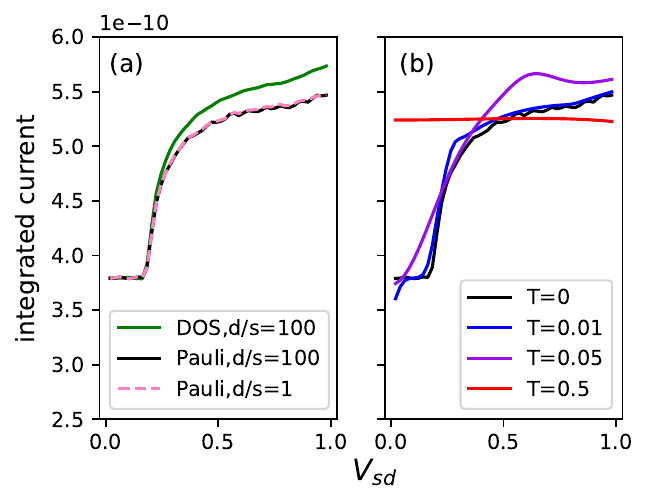}
\caption{\label{Fig:2dots_zba}
Integrated current (a) at $T=0$ comparing results from the DOS approach with $t_d/t_s=100$, the Pauli approach with $t_d/t_s=100$, and the Pauli approach with $t_d/t_s=1$ (multiplied by a factor of 2.25); and (b) Pauli calculations with $t_d/t_s=0$ comparing four temperatures.  The DOS calculation gives the true single-particle density of states as a function of frequency, while the Pauli calculations more accurately reflect what would be observed in an experiment and still show the kinetic-energy-driven zero-bias anomaly unique to the strongly correlated system.  Both panels use dot parameters $U=8,\ V=0.6,\ t_h=0.6$ and lead parameters $\mu_d=V+U/2,\ \mu_s=\mu_d+V_{sd},\ V_{sd}=0.6,\ t_s=0.0001$.
}
\end{figure}

Proceeding to our second question, does this change in the stability diagram mean that the SZBA is no longer visible in the integrated current as a function of bias voltage?  
Fig.\ \ref{Fig:2dots_zba}(a) shows the integrated current as a function of bias voltage in three different cases.
The green curve shows the result obtained using the DOS approach (the same as the green curve in Ref.\ \onlinecite{Wortis2022} Fig.\ 4(e)).  In this case, the integrated current is directly proportional to the DOS of an ensemble of two-site systems.
The black curve shows the result obtained using the Pauli approach with $t_d/t_s=100$.  
The zero bias anomaly is still clearly apparent.
Because the bright crescent of triplet transitions in region i12st enters at the same bias voltage in both cases, the width of the SZBA is consistent between the two methods. 
The only difference is a reduction in the height at higher biases, a direct result of the shadow in the i23st region which blunts the effect of the increase in current in the i12st region, although only slightly.

Next we asked the question of how these results would be affected by shifting from asymmetric to symmetric lead coupling.  
Fig.\ \ref{Fig:2dots_sd}(c) shows the stability diagram calculated also using the Pauli approach but here while $t_s$ is held fixed $t_d$ is reduced such that $t_d/t_s=1$.
With suppressed hopping amplitude to the dot, unsurprisingly current is reduced.
In addition the result shows symmetry not present in the $t_d/t_s=100$ case.
When coupling to the drain is equally strong as that to the drain, the occupation probabilities are set by where the energy differences $\Delta_{\alpha \to \beta}$ sit relative to the chemical potential of the source $\mu_s$, rather than to $\mu_d$.  
This means that more states have nonzero steady state occupation probabilities, as summarized in Appendix \ref{app:occup}.
For many of these additional states, current flow is not possible.
For example, in the i12s region initial state probability is equally divided between $\ket{1b\up}, \ \ket{1b\dn}$, and $\ket{2sa}$, but no transitions are accessible through the $\ket{2sa}$ state.
The result is an overall reduction in the current flow.
Nonetheless, despite the differences between Fig.\ \ref{Fig:2dots_sd} (b) and (c), and the underlying differences in transitions, the corresponding integrated current versus bias voltage shows the same pattern, as shown in the dashed pink curve in Fig.\ \ref{Fig:2dots_zba}(a).

\begin{figure}
\includegraphics[width=\columnwidth]{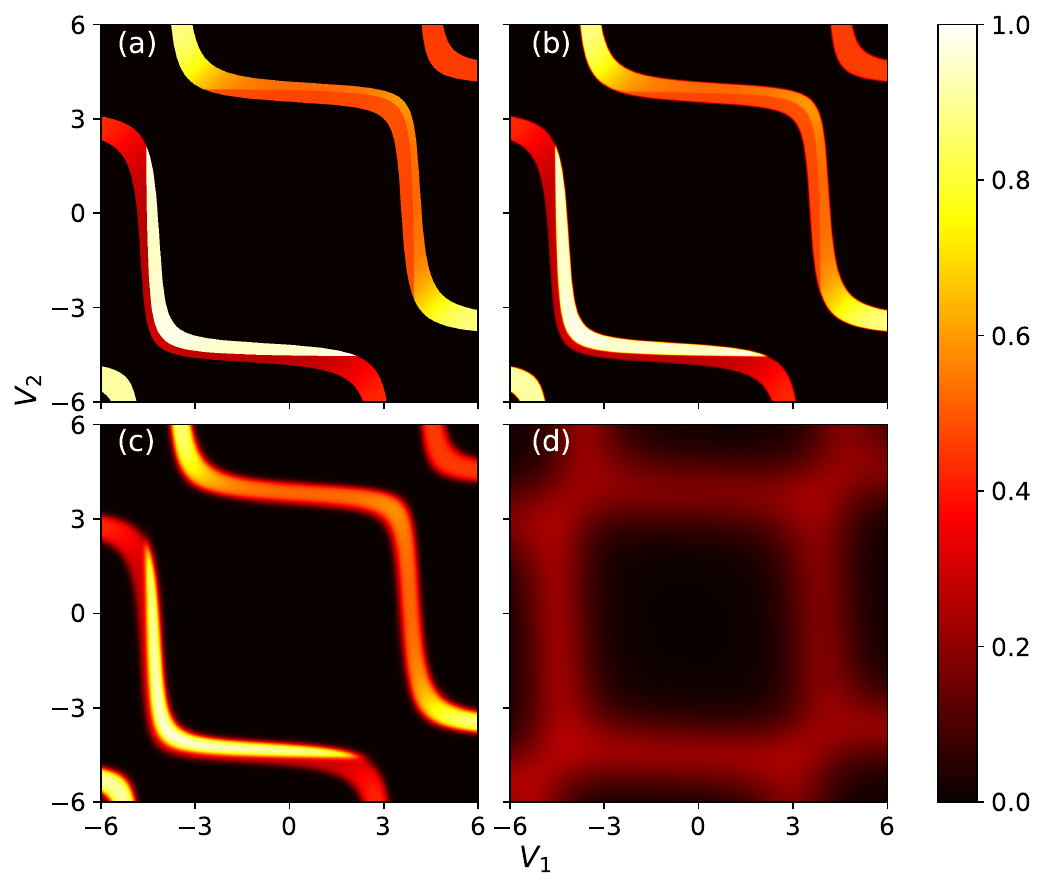}
\caption{\label{Fig:2dots_t}
Stability diagrams for a parallel-coupled double quantum dot calculated using the Pauli equation approach described in the text at temperatures (a) $T=0$, (b) $T=0.01$, (c) $T=0.05$, and (d) $T=0.5$.  Horizontal and vertical axes are the potential energy on the two dots, and the colour scale indicates the magnitude of the current.  All panels use dot parameters $U=8,\ V=0.6,\ t_h=0.6$ and lead parameters $\mu_d=V+U/2,\ \mu_s=\mu_d+V_{sd},\ V_{sd}=0.6,\ t_s=0.0001$ and $t_d/t_s=100$.
}
\end{figure}

Finally we looked at the changes brought about by an increase in temperature.
Fig.\ \ref{Fig:2dots_t} shows the stability diagram obtained using the Pauli method with $t_d/t_s=100$ and $T=0, 0.01$, 0.05, and 0.5 respectively, and Fig.\ \ref{Fig:2dots_zba}(b) shows the corresponding integrated current.
Fig.\ \ref{Fig:2dots_t}(b) is nearly indistinguishable from Fig.\ \ref{Fig:2dots_t}(a), just a slight blurring at the edges.
The blurred edges come from two sources.
First, current flow extends toward lower dot voltages (toward the lower left of each current region) due to thermal excitation of current-carrying states above the chemical potential.  
Second, current flow is reduced at the high dot voltage edge of each current region.
For example, near the upper right edge of the i12 region, where at zero temperature $P(1b\up)=P(1b\dn)=1/2$, at $T>0$ $P(2sa)$ becomes nonzero.
The energy for transitions from $\ket{2sa}$ to 3-particle states are far above $\mu_s$, so the result is a reduction in the flow of current.
At very low bias, this latter effect dominates, and there is a reduction in the integrated current as seen in Fig.\ \ref{Fig:2dots_zba}(b).
This very narrow anomaly also occurs in atomic-limit systems\cite{Wortis2011}(when no hopping is allowed between the dots), making it distinct physics from the broader and deeper kinetic-energy driven anomaly which is unique to strongly-correlated systems and the focus of this work.

In Fig.\ \ref{Fig:2dots_t}(b) while the top and bottom edges of each current region are blurred, the bright sliver in i12 and the shadow in i23 remain visible.
In Fig.\ \ref{Fig:2dots_t}(c), the higher temperature increases the width of the blurred edges.
In addition, the shadow (the i23st region) is lost, while the bright sliver (the i12st region) persists. 
The integrated current still shows a suppression around zero bias but without sharp features.
Finally, at $T=0.5$, all distinct features are lost and no sign of the SZBA remains in the integrated current.


\section{Conclusion}

The central result of this work is that a feature in the single-particle density of states which is unique to strongly correlated systems remains visible in the stability diagram of a parallel-coupled double quantum dot when the effect of lead coupling is included.
Whereas prior work\cite{Wortis2022} assumed current arose exclusively through transitions from a single ground state, here we calculate the steady-state occupation probabilities within a weak-coupling approach.  
This more accurate description results in an additional feature in the stability diagram.  Nonetheless, the physics of the two-site strongly-correlated zero-bias anomaly is evident both in the stability diagram and in the integrated current as a function of bias voltage for both asymmetric and symmetric lead coupling and over a nonzero temperature range.
Aspects of the physics described here is already apparent in published data.\cite{Chen2004,Hatano2013}
By establishing that signatures of the strongly-correlated zero-bias anomaly persist in the presence of lead coupling and characterizing the effect of temperature on these signatures, we hope this work will assist experimentalists in observing the strongly-correlated zero-bias anomaly in ensembles of two-site systems.

\section*{Acknowledgments}
We acknowledge support by the National Science and Engineering Research Council (NSERC) of Canada.

\appendix

\section{Coherence factors}
\label{app:cf}

The factors $C_{\alpha \beta}$ arise from the matrix element in Fermi's Golden Rule, Eqn.\ (\ref{eqn:fgr}).  Specifically, the part associated with the states of the dot.
When the number of particles in the state $\beta$ is one more than that in $\alpha$,
\begin{eqnarray}
C_{\alpha \beta} &=& \sum_{j,\sigma} | \bra{\beta} {\hat d}_{j\sigma}^{\dag} \ket{\alpha} |^2 
\end{eqnarray}
whereas when the number of particles in $\beta$ is one less than that in $\alpha$
\begin{eqnarray}
C_{\alpha \beta} &=& \sum_{j,\sigma} | \bra{\beta} {\hat d}_{j\sigma} \ket{\alpha} |^2.
\end{eqnarray}
Table \ref{tbl:c} summarizes the results.
Because $C_{\alpha \beta}$ depends on the magnitude squared of the matrix element and $\bra{\beta} {\hat d}_{j\sigma}^{\dag} \ket{\alpha} = \bra{\alpha} {\hat d}_{j\sigma} \ket{\beta}^*$, $C_{\alpha\beta}=C_{\beta\alpha}$. 

\begin{table}
\begin{tabular}{|r || c|c|c|  c| c|c|c|  c|c|c|c|} \hline 
& 0 & $1i\up$ & $1i\dn$ & $2sj$ & $2t\up$ & $2t0$ & $2t\dn$ 
	& $3i\up$ & $3i\dn$ & 4 \\ \hline \hline
0 	& . & 1 & 1 
	& . & . & . & . 
	& . & . & .  \\
$1i\up$ & 1 & . & .
	& $I_{1,2s}$ & 1 & 1/2 & . 
	& . & . & .  \\
$1i\dn$ & 1 & . & .
	& $I_{1,2s}$ & . & 1/2 & 1 
	& . & . & .  \\
$2sj$
	& . & $I_{2s,1}$ & $I_{2s,1}$ 
	& . & . & . & . 
	& $I_{2s,3}$ & $I_{2s,3}$ & . \\
$2t\up$ & . & 1 & . 
	& . & . & . & . 
	& 1 & . & . \\
$2t0$ & . & 1/2 & 1/2
	& . & . & . & . 
	& . & 1 & . \\
$2t\dn$ & . & . & 1 
	& . & . & . & . 
	& . & 1 & . \\
$3\up$ & . & . & . 
	& $I_{3,2s}$ & 1 & 1/2 & . 
	& . & . & 1 \\
$3\dn$ & . & . & .
	& $I_{3,2s}$ & . & 1/2 & 1 
	& . & . & 1 \\
4 & . & . & . 
	& . & . & . & . 
	& 1 & 1 & . \\
\hline
\end{tabular}
\caption{\label{tbl:c}
Values of the factors $C_{\alpha \beta}$.  $\alpha$ values are listed on the left and $\beta$ values across the top.  Entries not equal to 1 or 1/2 are written $I_{\alpha \beta}$ with these defined in Eqns.\ (\ref{eqn:Iab12}) and (\ref{eqn:Iab23}).  
}
\end{table}

Using the notation for the eigenstates in terms of Fock states
\begin{eqnarray}
\ket{0} &=& \ket{00} \\
\ket{1i\sigma} 
	&=& a_{1i\sigma,\sigma 0} \ket{\sigma 0}
		+ a_{1i\sigma, 0 \sigma} \ket{0 \sigma} 
		\nonumber \\
		& & i=a,b;\ \sigma=\up,\dn \\
\ket{2sj} 
	&=& a_{2sj,s} \ket{s} + a_{2sj,20} \ket{20} + a_{2sj,02} \ket{02} 
	\nonumber \\
	& & j=a,b,c; \ \ket{s} \ = \ (\ket{\up \dn} - \ket{\dn \up})/\sqrt{2} \\
\ket{2t0} &=& (\ket{\up \dn} + \ket{\dn \up})/\sqrt{2} \\
\ket{2t\up} &=& \ket{\up \up};\ 
	\ket{2t\dn} \ = \ \ket{\dn \dn} \\
\ket{3i\sigma} 
	&=& a_{3i\sigma,2\sigma} \ket{2 \sigma} 
		+ a_{3i\sigma,\sigma 2} \ket{\sigma 2} \\
\ket{4} &=& \ket{22}		
\end{eqnarray}
we can write the two non-trivial coherence factors
\begin{eqnarray} 
I_{1,2s} 
&=& I_{2s,1} \ = \ C_{2sj,1i\sigma} \ = \ C_{1i\sigma,2sj}  \nonumber \\
&=&| a_{2sj,20}^* a_{1i\sigma,\sigma 0} + (a_{2sj,s}^*/\sqrt{2}) a_{1i\sigma,0 \sigma} |^2 \nonumber \\ & & 
	+ | a_{2sj,02}^* a_{1i\sigma,0 \sigma} + (a_{2sj,s}^*/\sqrt{2}) a_{1i\sigma,\sigma 0} |^2
\label{eqn:Iab12}
\\
I_{2s,3} &=& I_{3,2s} \ = \ C_{2sj,3i\sigma} \ = \ C_{3i\sigma,2sj} \nonumber \\
&=& | a_{3i\sigma,2\sigma}^* (a_{2sj,s}/\sqrt{2}) - a_{3i\sigma,\sigma 2}^* a_{2sj,02} |^2 \nonumber \\ & & 
	+ | a_{3i\sigma,\sigma 2}^* (a_{2sj,s}/\sqrt{2}) - a_{3i\sigma,2\sigma}^* a_{2sj,20} |^2 
\label{eqn:Iab23}
\end{eqnarray}

\section{Occupation probabilities}
\label{app:occup}

\begin{figure}
\includegraphics[width=\columnwidth]{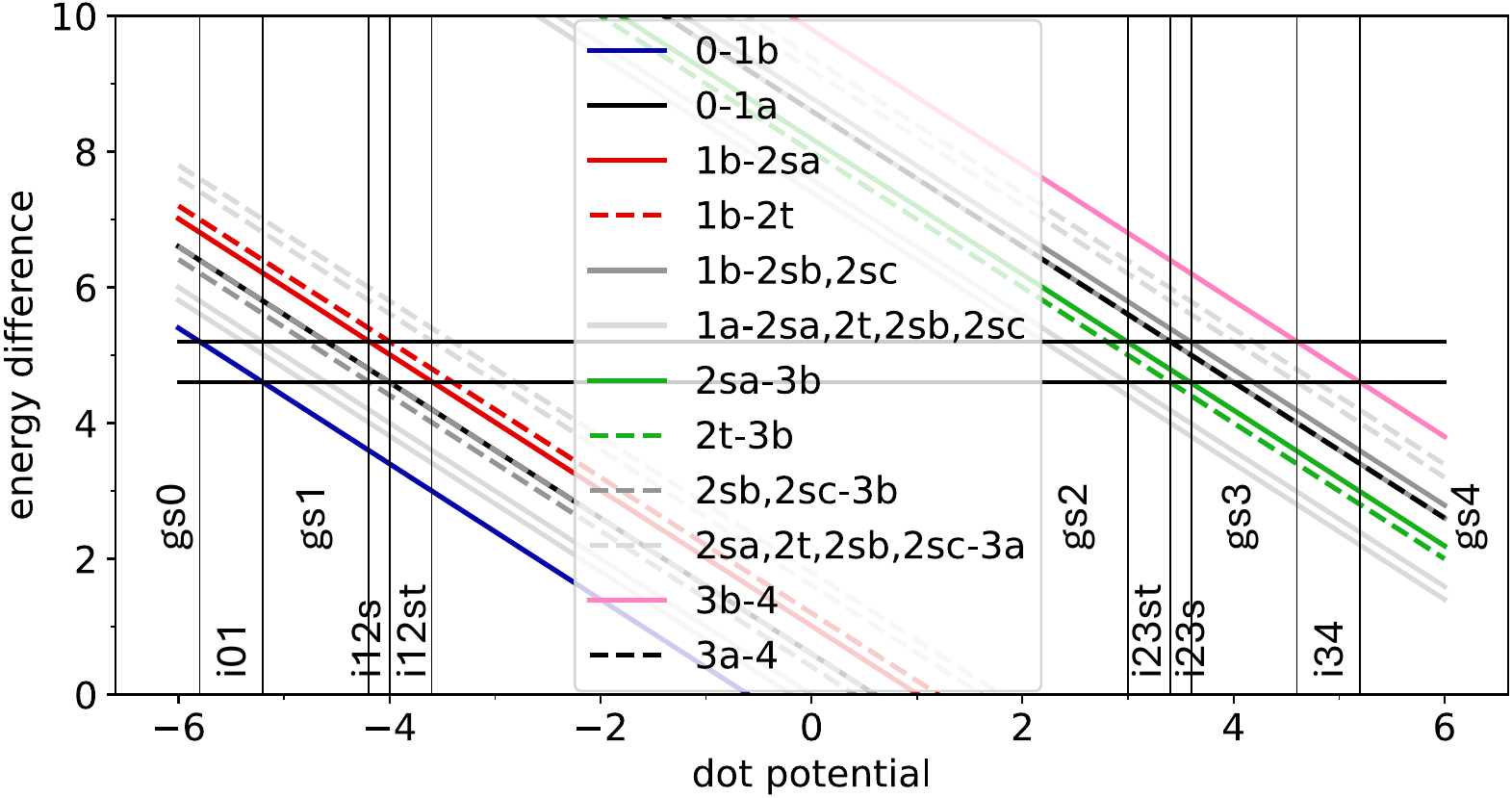}
\caption{\label{fig:energydifferences}Energy differences $\Delta_{\alpha \to \beta}=E_{\beta}-E_{\alpha}$ corresponding to all single particle transitions plotted as a function of the dot potential along the diagonal in the stability diagram, $V_1=V_2$.  Also shown are the chemical potentials of the source and drain.
}
\end{figure}

\begin{table*}
\begin{tabular}{|c|l|r|r|} \hline
& DOS approach & \multicolumn{2}{|c|}{Pauli approach} \\ 
& ground state & \multicolumn{2}{|c|}{occupation probabilities} \\ \hline
& & $t_d/t_s=100$ & $t_d/t_s=1$ \\ \hline
i01 & $\ket{0}$ 		& $P(0)=1$ 
			& $P(0),P(1b\up),P(1b\dn)=1/3$ \\
i12s & $\ket{1b\up}$ or $\ket{1b\dn}$ & $P(1b\up),P(1b\dn)=1/2$ 
			& $P(1b\up),P(1b\dn),P(2sa)=1/3$ \\
i12st & $\ket{1b\up}$ or $\ket{1b\dn}$ & $P(1b\up),P(1b\dn)=1/2$ 
			& $P(1b\up),P(1b\dn),P(2sa),P(2t,+1),P(2t,0),P(2t,-1)=1/6$ \\
i23st & $\ket{2sa}$ 	& $P(2sa),P(2t,+1),P(2t,0),P(2t,-1)=1/4$ 
			& $P(2sa),P(2t,+1),P(2t,0),P(2t,-1),P(3b\up),P(3b\dn)=1/6$ \\
i23s & $\ket{2sa}$ 	& $P(2sa)=1$ 
			& $P(2sa),P(3b\up),P(3b\dn)=1/3$ \\
i34 & $\ket{3b\up}$ or $\ket{3b\dn}$ & $P(3b\up),P(3b\dn)=1/2$ 
			& $P(3b\up),P(3b\dn),P(4)=1/4$ \\ \hline
\end{tabular}
\caption{\label{tbl:2dotregions}}
\end{table*}

In the DOS approach, all transitions occur from the ground state as determined by the minimum grand potential using $\mu_d$ as the chemical potential of the dot.  
Table \ref{tbl:2dotregions} summarizes the resulting ground state in each region with current in the stability diagram.

In the Pauli approach, transitions originate from a set of states with nonzero probability of being occupied.
These states are determined by (i) the energy differences between states $\Delta_{\alpha \to \beta}\equiv E_{\beta} - E_{\alpha}$ and (ii) a chemical potential $\mu$.
When $t_d/t_s>>1$, in steady state, when electrons enter the dot they immediately exit through the drain.  Therefore, in this limit the relevant chemical potential is $\mu_d$.
In contrast, when $t_d/t_s=1$ occupancy of states in the bias window is not negligible and $\mu_s$ becomes the relevant chemical potential.
Fig.\ \ref{fig:energydifferences} shows the energy differences for all single-particle transitions as a function of dot potential along the diagonal line in the stability diagram for which $V_1=V_2$.

Probability is equally distributed among all accessible states with maximum particle number.  
The state $\gamma$ is accessible if there is a sequence of single-particle transitions $\alpha \to \beta$ from the vacuum to the state $\gamma$ such that $\Delta_{\alpha \to \beta}<\mu$ in all cases.
$\gamma$ has maximum particle number if in addition $\mu<\Delta_{\gamma \to \delta}$ for all states $\delta$ with higher particle number than $\gamma$.

Consider for example the region labeled gs1 in the case $t_d/t_s=100$.  Here $\Delta_{0 \to 1b}<\mu_d$, so the states $1b\sigma$ are accessible.  The only other transitions with energies below $\mu_d$ in this region are $1a \to 2sa$ and $1a \to 2t$.  However, $\Delta_{0 \to 1a}>\mu_d$ so $1a$ and hence also $2sa$ and $2t$ are not accessible.  Moreover, there are no transitions out of $1b$ which fall below $\mu_d$, hence the two states $1b\up$ and $1b\dn$ have maximum particle number.  Distributing probability equally between them, $P(1b\up)=P(1b\dn)=1/2$. 
In this case these two states have equal energy, but in the gs2 region the states $2sa$ and $2t\tau$ are all accessible with maximum particle number.  They therefore all have equal probabilities even though they differ in energy.
The probability values in all current regions are summarized in Table \ref{tbl:2dotregions} for $t_d/t_s$ both large and small.


%

\end{document}